# Master Sintering Curve with dissimilar grain growth trajectories: A case study on MgAl$_2$O$_4$

Gabriel Kerbart[1], Charles Manière[1], Christelle Harnois[1], Sylvain Marinel[1]

1 Normandie Univ, ENSICAEN, UNICAEN, CNRS, CRISMAT, 14000 Caen, France

**Abstract**

Sintering is a key step in the processing of high performance ceramics. Both the density and the grain size play a crucial role on the ceramic sintering kinetics and the final material properties. The master sintering curve (MSC) is a well-known tool for exploring sintering models kinetics. However, the conventional MSC theory assumes a unique sintering trajectory, while our study on MgAl$_2$O$_4$ spinel shows dissimilar growth response. Park's MSC theory has been applied and compared with the conventional MSC approach for obtaining the activation energy with and without dissimilar grain growth trajectories.

**Keywords**

Sintering; Master Sintering Curve; Grain growth; Transparent Ceramics; MgAl$_2$O$_4$; Modeling

**Graphical abstract**

**Conventional MSC assumption: Unique trajectory**

$$\Theta = \int_0^t \frac{1}{T} \exp\left(\frac{-Q}{RT}\right) dt$$

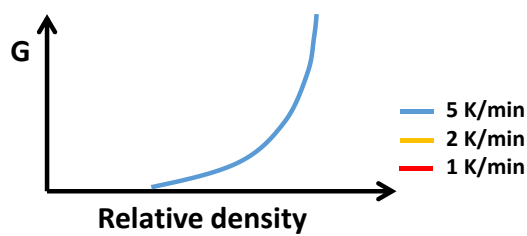

**MSC with independent grain growth: ≠ trajectories allowed**

$$\Theta = \int_0^t \left(\frac{G_0}{G}\right)^4 \frac{1}{T} \exp\left(\frac{-Q}{RT}\right) dt$$

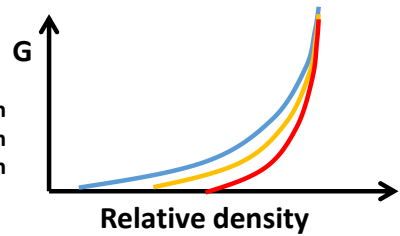

# I- Introduction

A well-known tool of ceramist to study the sintering is the Master Sintering Curve concept (MSC) initially developed by Su and Johnson[1]. This tool is used to determine the activation energy of a solid state sintering model under the assumption of a single dominant diffusion mechanism[2–4]. MSC theory implies the integration of the sintering model with, on one hand, the diffusion Arrhenius temperature dependence with time and, on the other hand, all others sintering terms, grain size, porosity function, relative density and constants. For the latter, the grain size (G) which is typically time and temperature dependent is assumed to depend only on the porosity (θ). Consequently, the so-called "sintering trajectory" G(θ) is assumed to be unique for all heating rates, which is the conventional MSC concept. This assumption allows an easy identification of the activation energy but prevents from any sintering optimization. The final stage effects on MSC were detected and studied in various ceramics and metals[5–7]. For instance, a sintering optimization method like the "two stages sintering" is precisely based on dissimilar sintering trajectories[8]. Park et al[9] generalized the concept of MSC including both densification and grain growth evolution. This generalized conceptualization of MSC allows a better understanding of the sintering trajectory dissimilarities which are the basis of sintering optimizations. In the literature, the final stage phenomenon is also considered *via* transition of mechanism and/or activation energy in the MSC[10]. The transition may be used to correct the MSC sintering response of two-step sintering[10]. We will show in this study that both grain growth and mechanism transition may be considered.

This paper focuses on $MgAl_2O_4$ spinel. This material attracts much interest from the scientific community to industries for transparent polycrystalline ceramics applications like laser host, jewelry, spacecraft windows, IR windows and dome for military applications[11]. The isotropic structure of this material is a great advantage for producing transparent ceramics as it is not sensitive to the detrimental grain size effect on the optical transmission coefficient (due to the birefringence phenomenon)[12,13]. However, to ensure high mechanical properties and the absence of intra-grain pores, a careful control of grain growth is required[14]. For $MgAl_2O_4$ spinel, we have previously shown that the grain growth law changes from surface diffusion for high porosity content, to lattice diffusion for low porosity content[15]. The resulting grain growth model, inspired by Olevsky et al.[11] and Zhao and Harmer[17], therefore includes a transition mechanism with a porosity dependence. This advanced model allows reporting dissimilar sintering trajectories if needed. In this paper, the MSC tool is used via the conventional method (unique sintering trajectory) and with the Park's equations, which may involve different sintering trajectories. The impact of the grain growth during the final stage of the sintering process will be discussed accordingly to these two distinct models.

# II- Theory and calculation

The MSC theory predicts a unique curve plotting the relative density vs the following term:

$$\Theta = \int_0^t \frac{1}{T} \exp\left(\frac{-Q}{RT}\right) dt \qquad (1)$$

with, T the absolute temperature, t the time, R the gas constant and Q the activation energy. For dissimilar grain growth trajectories, Park's model[9] gives:

$$\Theta = \int_0^t \left(\frac{G_0}{G}\right)^w \frac{1}{T} \exp\left(\frac{-Q}{RT}\right) dt \tag{2}$$

with, $G_0$ the initial grain size and G the actual grain size, w the sintering mechanism exponent of 4 and 3 for grain boundary and lattice diffusion respectively. In this study, we assume grain boundary diffusion dominates.

The latter MSC approach considers the independent grain growth influence assuming no pore trapped gas influence on the final stage (high solubility of gas in solid phase or grain boundaries). The use of Park's model implies the knowledge of grain size throughout the whole sintering process. Olevsky's grain growth model[16] was chosen for $MgAl_2O_4$ to predict the grain growth affected by porosity:

$$\dot{G} = \frac{K(T)}{G^p}\left(\frac{\theta_c}{\theta + \theta_c}\right)^n \tag{3}$$

with p, n mechanism exponents, $\theta$ the porosity, $\theta_c$ the critical porosity and $K(T)$ an Arrhenius term.

In a previous study on the same powder[15], the grain growth was determined experimentally focusing on the transition region from the onset of grain growth at low porosity to the grain growth behavior at high densification. We evidence a complex transition mechanism from high porosity to full densification requiring a change of the exponent *p* and the porosity function of Olevsky's model. This advanced grain growth model is of particular interest to use with MSC as a careful modeling of grain growth is required in the whole porosity range.

### III – Experiment and method

The samples of this study were produced by cold isostatic pressing (CIP) at 300 MPa of a commercial Baikowski S30CR spinel powder. To avoid disturbance from binder or anisotropic sintering shrinkage, the green specimens were prepared without binder or previous uniaxial pressing (UP). This powder has a purity over 99%, a specific surface area between 25 and 28 m²/g (BET) and a d50 between 0.15 and 0.3 µm. The samples were sintered up to 1773 K with three different heating ramps of 1, 2 and 5 K/min.

For this purpose, we used a dilatometer (setsys 16/18, SETARAM, France) to record the vertical displacement of an alumina rod in contact with the surface of the samples. The final density of these samples was determined by Archimedes's method. Assuming the shrinkage is isotropic, we can calculate the density versus time during the overall sintering cycle (see Fig. 1). As expected, the density increases with lower heating rates. This sintering curves (Fig. 1) recorded on materials prepared *via* pure CIP without binder are very closed to those reported by Yalamaç results (UP + CIP) [18]. Compared to our results, and using UP+CIP, Talimian *et al*[19] found shrinkage curves slightly shifted towards lower temperature, whereas Benameur *et al*[20] reported shrinkage curves slightly shifted towards higher temperature (slip casting). These literature data show the slight impact of the powder shaping method on the sintering temperature. The shrinkage curves data are sufficient when using the Su and Johnson conventional MSC method. However, the use of Park's MSC model requires the knowledge of the grain size dependence with temperature/time. The method for this purpose is detailed hereafter.

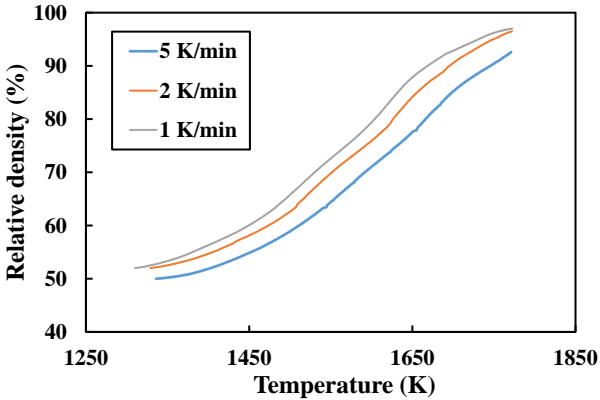

Figure 1 Relative density versus temperature for each heating ramp.

### IV – Results and discussion

#### 1. Grain growth modeling

The modeling of grain growth, during sintering, is based on the Olevsky's equation (Eq.3). This model allows predicting the grain growth for the different heating rates, which is needed for the MSC Park's model.

We show on Fig. 2. (a) the results of the grain growth modeling applied to the three dilatometric ramping cycles. As expected, lower heating rates led to bigger grain size. The development of this model (from a previous study[15]) is based on the analysis of twelve samples, sintered at three different temperatures (1673 K, 1773 K, and 1873 K) for four different dwell times (0h, 1h, 2h, 4h). The determination of the grain size and the porosity of these samples allows for the determination of the Eq.3 model parameters. As a result, the grain size dependence with density can be modeled as shown on figure 2 (b). The model can be applied to MSC cycles using the dilatometric data for the porosity function of Eq.3

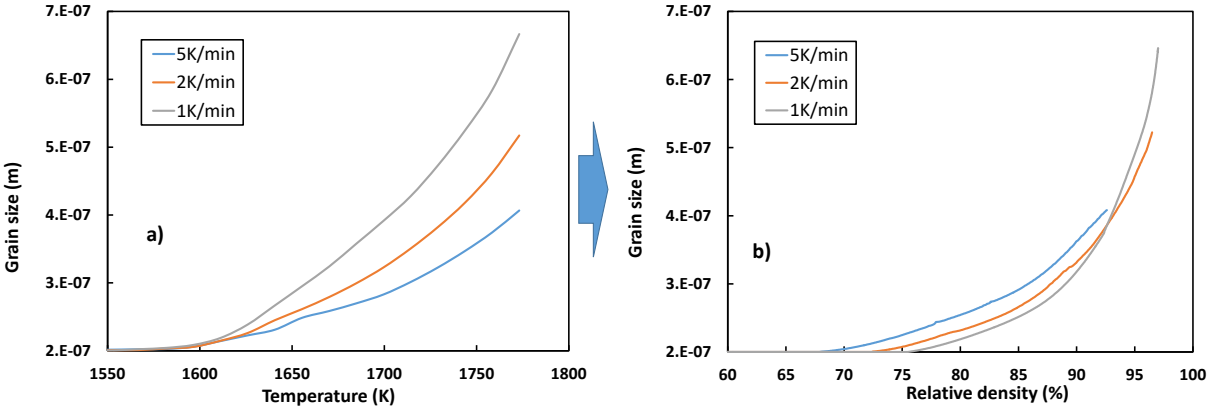

Figure 2 MSC Grain size evolution during sintering and corresponding sintering trajectory curves.

## 2. Master Sintering Curve with and without grain growth

We show on Fig. 3. both MSC, realized respectively with Su and Johnson model (Fig. 3. a), and Park's (Fig. 3. b). On each graph, it is shown the relative density versus the logarithm of the Θ function (see Eq. 1. and Eq. 2.) while in the insert is shown the mean of residual squares versus the tested activation energy. The use of Su and Johnson and Park's MSC models has conducted to a close estimation of the activation energy of 450 and 485 kJ/mol, respectively. The two MSC have relatively close activation energies with a difference of 35 kJ/mol. Nevertheless, an apparent difference between these two MSC models is the slightly higher mean of residual squares for Park's model. This explains the weaker superposition of the curve for Park's approach.

The only difference between the two models is the dissimilar grain growth trajectory which is taken into account for Park's model. As shown on Fig. 2. grain growth starts at 1600 K for each sample, and this temperature corresponds to different densities due to the different heating rates. Therefore, the sintering trajectories cannot be considered unique in the range of (65 - 90 %) of density, as clearly shown figure 2(b). There is a noticeable effect when the relative density is above 80 %, on the figure 3 (b), it can be seen that the superposition of the different curves is much better, with a very low mean of residual squares. This relative density range corresponds to the domain where grain growth is taking place (see fig.2 b). If we split Park's MSC analysis in two zones "with/without grain growth" in Fig. 3. c and Fig. 3. d, a better superposition is observed with different activation energies (400 kJ/mole and 700 kJ/mole at low and high densities respectively). This may indicate a change in the densification mechanism when grain growth starts. This two zones MSC study maybe be useful for comparing the different activation energies for each sintering stages[21].

In the literature, the MSC activation energies for the same powder grade are ranging from 750 to 950 kJ/mole for UP + CIP[18] and slip casting[20] samples respectively. Our values are closer to the 530 kJ/mole obtained by Talimian *et al*[19] on ground powder. Our green specimen preparation without binder, and shaped by pure CIP, seems to mimic the behavior of the ground powder prepared by Talimian et al.

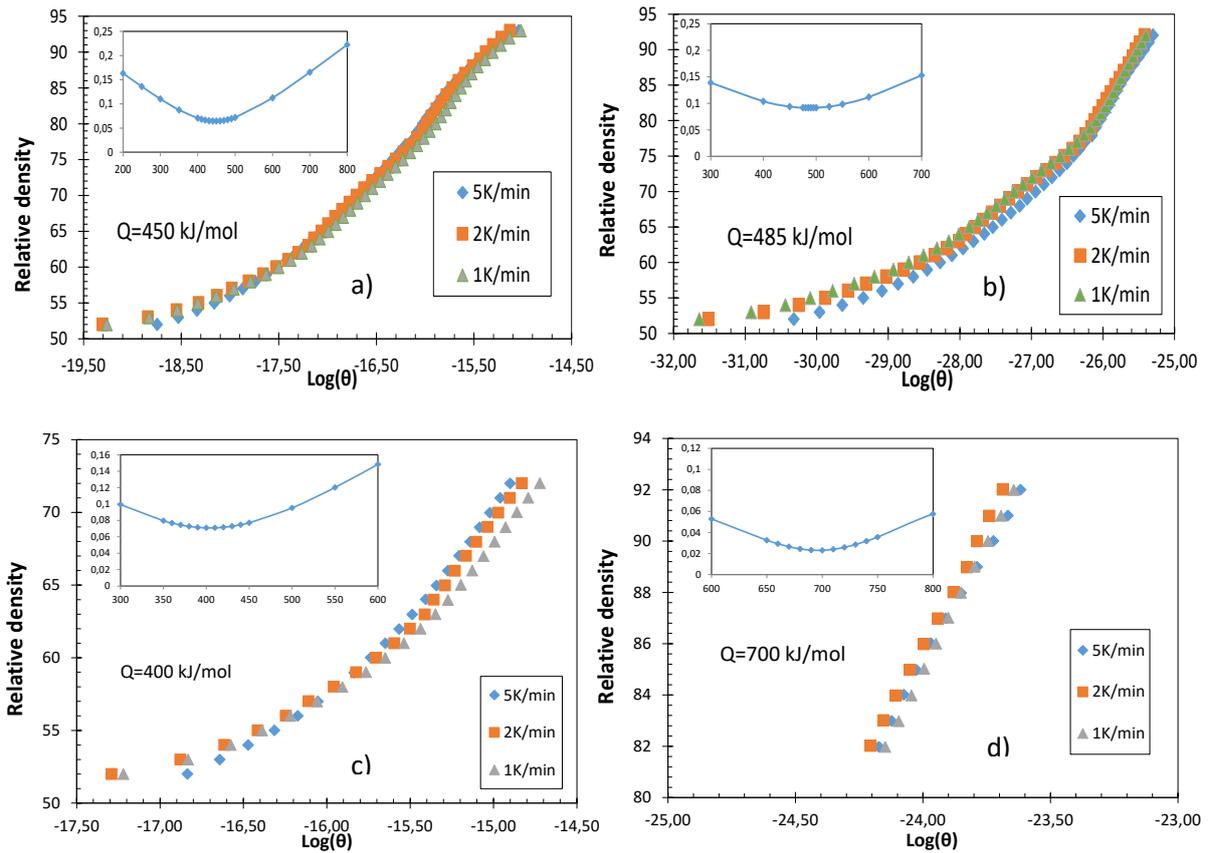

*Figure 3 a) Su and Johnson model for spinel sintering. b) Park model for spinel sintering. c) Park model at low density. d) Park model at high density.*

## V- Conclusion

Final stage sintering grain growth is a complex phenomenon influenced by the porosity and transition mechanism. Gain growth extends the diffusion distance and then decrease the sintering kinetics at the final stage. Controlling the densification/grain growth competition at the final stage is one of the key issues for obtaining high-performance ceramics and the sintering modeling can help predict such phenomena. The MSC is a method typically employed to predict the sintering model parameters such as the activation energy. However, the conventional method assumes a unique sintering trajectory instead of considering the final stage grain growth effect as a temperature-dependent separated variable. Based on Park's approach, the MSC has been investigated taking into account the final stage independent grain growth. In a previous study, the spinel grain growth behavior has been identified in a large temperature range and for the same preparation/sintering conditions. This has allowed us to test Park's MSC approach without having to conduct an extensive study to determine the grain growth of each curves. This study gives the following conclusions.

1    The sintering trajectories obtained from the densification curves and the grain growth model is not converged into a single curve as supposed by conventional MSC approach but shows noticeable dissimilarities.

2   The sintering trajectory dissimilarity implies a higher value of activation energy compared to the conventional MSC.
3   Park's MSC was also investigated via the two zones MSC consisting of estimating separately the activation energy in the initial/intermediate stage and in the final stage. Significant differences in the activation energy confirm the lowering of the sintering kinetics by grain growth.
4   The preparation of green specimens by CIP without binder seems to favor lower activation energy compared to the literature results.

**Acknowledgement**

The authors would like to thank Baikowski for the powder, and the French ministry of research for the funding of the thesis. The authors are also thankful to Jerome Lecourt and Christelle Bilot for their precious help in conducting the experiments.

**Figure Captions**

Fig. 1. Relative density versus temperature for each heating ramp.

Fig. 2. MSC Grain size evolution during sintering and corresponding sintering trajectory curves.

Fig. 3. a) Su and Johnson model for spinel sintering. b) Park model for spinel sintering. c) Park model at low density. d) Park model at high density.